%% file: ZS-ajp-rev.tex
\documentstyle[aps,prb,twocolumn,epsfig,amsmath,amssymb]{revtex}
\input epsf.sty

\begin{document}

\twocolumn[\hsize\textwidth
\columnwidth\hsize\csname
@twocolumnfalse\endcsname

\title{Watching a drunkard for ten nights: A study of distributions of
variances}
\author{R. K. P. Zia and B. Schmittmann}
\address{Center for Stochastic Processes in Science and Engineering,\\
Physics Department, Virginia Polytechnic Institute and State University,\\
Blacksburg, VA, 24061-0435 USA}
\date{\today}
\maketitle

\begin{abstract}
For any physical observable in statistical systems, the most frequently
studied quantities are its average and standard deviation. Yet, its full 
{\em distribution} often carries extremely interesting information and can
be invoked to put any surprising properties of the individual moments into
perspective. As an example, we consider a problem concerning simple random
walks which was posed in a recent text. When a drunk is observed over $L$
nights, taking $N$ steps per night, and the number of steps to the right is
recorded for each night, an average and a variance based on these data can
be computed. When the variance is used to estimate $p $, the probability for
the drunk to step right, {\em complex} values for $p $ are frequently found.
To put such obviously nonsensical results into context, we study the full
probability distribution for the variance of the data string. We discuss the
connection of our results to the problem of data binning and provide two
other brief examples to demonstrate the importance of full distributions.
\end{abstract}
\vspace{0.3cm}
]

\narrowtext

\section{Introduction}

Many properties of the random walk are well-known \cite{Hughes} and form an
important part of good texts in statistical mechanics, at either the
undergraduate or graduate level \cite{texts}. Despite its ``age,'' \cite%
{Pearson} this problem continues to present new and interesting puzzles,
depending on the questions asked of the walker. Many recent examples of such
puzzles, some of which remain unsolved, involve the issue of {\em full
distributions} (as opposed to just the averages) of certain quantities.
Providing an exhaustive list is beyond the scope of this paper. Instead, in
the concluding section, we discuss several examples to illustrate both the
value and the excitement in such studies. Our main interest lies in a
distribution rarely discussed in texts, in connection with a problem posed
in the manuscript of an undergraduate text book on statistical mechanics,
co-authored by H. Gould and J. Tobochnik (GT) \cite{GT}.

Let us first quote the question which motivated this paper, namely, the
first two parts of problem 3.37 in this manuscript.

\begin{quote}
A random walker is observed to take a total of $N$ steps, $n$ of which are
to the right.

{\em (a)} Suppose that a curious observer finds that on ten successive
nights the walker takes $N=20$ steps and that the values of $n$ are given
successively by 14, 13, 11, 12, 11, 12, 16, 16, 14, 8. Compute $\bar{n}$, $%
\overline{n^2}$, and $\sigma _n$. Use this information to estimate $p$. If
your reasoning gives different values for $p$, which estimate is likely to
be the most accurate?

{\em (b)} Suppose that on another ten successive nights the same walker
takes $N=100$ steps and that the values of $n$ are given by 58, 69, 71, 58,
63, 53, 64, 66, 65, 50. Compute the same quantities as in part (a) and
estimate $p$. ... Explain your results.
\end{quote}

To help the reader with context and notation, we add that this problem is at
the end of a section about a simple random walk in one dimension, stepping
either to the right or left, with probability $p$ and $1-p$, respectively.
The standard results for the average number of right steps, $Np$, and the
associated variance, $Np\left( 1-p\right) $, were derived. We will refer to
these expressions as the ``true average'' and the ``true variance,'' denoted
by 
\begin{equation}
n_{av}\equiv Np\quad \text{and}\quad v\equiv Np\left( 1-p\right)
\label{true-stuff}
\end{equation}
respectively. They would result if we were to observe the walker for very
(ideally, infinitely) many nights.

In the problem, however, $\bar{n}$ and $\sigma _{n}$ denote the average and
the standard deviation (i.e., $\sqrt{\overline{n^{2}}-\bar{n}^{2}}$)
computed from observations covering only a relatively {\em small} number
(ten) of nights. So, estimates for $p$ are to be made from either the
equation for the average 
\begin{equation}
\bar{n}=Np_{av}  \label{nbar}
\end{equation}
or one for the standard deviation 
\begin{equation}
\sigma _{n}=\sqrt{Np_{\sigma }\left( 1-p_{\sigma }\right) }\,\,.  \label{sn}
\end{equation}
The authors pointed out \cite{private} that, since the second method
involves a quadratic equation for $p$, there will be {\em two }solutions
(indeed, symmetric around{\em \ }$p=1/2$), so that this route cannot produce
a unique answer by itself.

More interestingly, if straightforward computations are carried out, it is
even more trivial to answer the question concerning which estimate is more
``accurate'': $p_{\sigma }$ is {\em complex} for both (a) and (b)! When we
trace the origins of this remarkable result, we find that, though the true
variance $v$ never exceeds $N/4$, the data string of a {\em %
particular} night can easily exceed this bound. One of the ``worst case''
scenarios -- focusing again on 10 nights -- occurs when the drunk takes $N$
steps to the right for $10p$ nights and $N$ steps to the left for $10\left(
1-p\right) $ nights. For example, instead of the first string given in part
(a) of the problem, we have (assuming $p=0.6$) 20,20,0,0,20,0,20,20,0,20.
Though such a string even provides the exact underlying $p$, it leads to $%
\sigma _{\max }^{2}=N^{2}p\left( 1-p\right) $. Note that $\sigma _{\max
}^{2}\,$ comes with even a {\em wrong power} of $N$, so that, for
sufficiently large $N,$ it will always exceed the largest possible true
variance, $\max_{p\in \lbrack 0,1]}Np\left( 1-p\right) =N/4$, regardless of
the underlying $p$! At the other extreme, if the same $n$ is observed on 
{\em every} night, then the absolute minimum $\sigma _{\min }=0$ is
achieved. Though this $\sigma _{\min }$ leads to a {\em real} $p_{\sigma }$,
the result ($p_{\sigma }=0$ or $1$) is clearly also unreasonable. Although
we can easily compute the expectation values of $\sigma _{n}$ and 
$p_{\sigma}$, a natural question arises: 
how likely is it that $\sigma _{n}^{2}$
exceeds the absolute bound of $N/4$? Similarly, we could ask for the
likelihood that $p_{\sigma }$ comes within, say, 5\% of the underlying $p$.
Thus, we are led to study the full distribution of variances in the
observations of a drunkard.

The rest of this paper is organized as follows. The next section is devoted
to the detailed analysis of the issues at hand. Section III is provided as a
convenience, for those who wish to skip the details for now, read a summary
of our conclusions, and see how they apply to the specific problem here. In
a final section, we make brief suggestions for expanding discussions of
distributions in typical courses on statistical mechanics.

\section{Distributions of averages and variances}

To motivate the study of distributions, let us use the language of human
behaviorists to define an ``ensemble'' of many {\em identical} drunkards
(say, $M$) residing in different cities. The probability of stepping to the
right is $p$, for {\em all} drunkards in {\em all} nights. Each is observed
for $L$ nights, $N$ total steps each night. The number of steps to the
right, $n$, differs each night and for different walkers. So, the entire
data set can be summarized by $M$ strings of $L$ numbers: 
\begin{equation}
n_{\alpha ,i}\,\,,\quad \alpha \in \left[ 1,M\right] ;\,i\in \left[ 1,L%
\right] \,\,,  \label{nai}
\end{equation}
with 
\[
0\leq n_{\alpha ,i}\leq N\,\,. 
\]
Alternatively, we can imagine the string of numbers in GT's text being
generated randomly (with the {\em same }$p$) each time the URL is accessed.
After $M$ readings, we would have $M$ different strings. So, for the two
cases stated in the problem above, $M$ is just $1$ and $L=10$, while $N=20$
and $100$.

From this big data set, we can construct $M$ averages and variances, each
generated from the data set for one of the drunks: 
\begin{equation}
\bar{n}_{\alpha }\equiv \frac{1}{L}\sum_{i}n_{\alpha ,i}\,,\,\,V_{\alpha
}\equiv \frac{1}{L}\sum_{i}n_{\alpha ,i}^{2}-\left( \bar{n}_{\alpha }\right)
^{2}\,\,.  \label{nV}
\end{equation}%
Needless to say, $0\leq \bar{n}_{\alpha }\leq N$ and $0\leq V_{\alpha }\leq
N^{2}/4$. (Note the $N^{2}$ in the last bound, which comes from {\em the
worst} case scenario with only right steps for exactly half the nights!)
Clearly, $\bar{n}_{\alpha }$ and $V_{\alpha }$ are still random variables,
with respect to the ensemble of drunks. If we make histograms from the two
sets of $M$ numbers, denoted by $H\left( \bar{n}\right) $ and $H\left(
V\right) $, we get a glimpse of the full distribution of their possible
values. The first is easy, being related to the binomial distribution, and
approaches a Gaussian for large $N$. Nevertheless, there are some subtleties
to which we will alert the reader. The second will give us an idea of how
often we can expect a data string to produce a variance which exceeds the
maximum possible value, for any realizable $p$, namely, $N/4$.

\subsection{Statistics of averages}

For completeness, let us remind the reader that, if we focus on the
statistics of a single drunkard taking $N$ steps on a single night, then the
probability that he takes $n$ right steps is just the binomial distribution: 
\begin{equation}
P\left( n\right) =%
{N \choose n}
p^n\left( 1-p\right) ^{N-n}\,\,.  \label{Pn}
\end{equation}
If he is observed for $L$ nights (for a total of $LN$ steps), then the
probability that he took a {\em total} of $m$ steps to the right is clearly
also a binomial: $%
{LN \choose m}%
p^m\left( 1-p\right) ^{LN-m}.$ But, notice that $\bar{n},$ the average over
the $L$ nights of right steps, is precisely $m/L$. Thus, we may conclude
immediately that the probability distribution for the {\em average} number
of right steps is $\,\,$ 
\begin{equation}
{\cal P}\left( \bar{n}\right) =%
{LN \choose L\bar{n}}
p^{L\bar{n}}\left( 1-p\right) ^{L(N-\bar{n})}\,\,.
\end{equation}
If we take data for $M$ drunkards and compile a histogram for the set of $%
\bar{n}_\alpha $, we should find 
\begin{equation}
H\left( \bar{n}\right) \rightarrow M{\cal P}\left( \bar{n}\right)
\end{equation}
as $M\rightarrow \infty $.

This distribution also tells us how reliable our estimate for $p$ is
when we have data for only one drunkard (as posed in the text problem). In
particular, as we can guess intuitively, although we are {\em most likely}
to get the right $p$ by just dividing $\bar{n}$ by $N$ (i.e., ${\cal P}%
\left( \bar{n}\right) $ peaks at $Np$), the chances we are off can be
estimated through the standard deviation in ${\cal P}\left( \bar{n}\right) $%
, namely, $Np\left( 1-p\right) /L$.

Finally, if $N$ is large, the binomial $P\left( n\right) $ is
indistinguishable from a Gaussian: 
\begin{equation}
\tilde{P}\left( n\right) =\frac{1}{\sqrt{2\pi v}}\exp \left[ -\frac{\left(
n-n_{av}\right) ^{2}}{2v}\right] \,\,.  \label{PGn}
\end{equation}%
Since convolutions of Gaussians form a Gaussian, the distribution of the
averages is also a Gaussian, with variance $v/L.\,$In other words, we should
find  
\begin{equation}
\frac{H\left( \bar{n}\right) }{M}%
\mathrel{\mathop{\rightarrow }\limits_{M\rightarrow \infty }}%
{\cal \tilde{P}}\left( \bar{n};n_{av},v,L\right) =\sqrt{\frac{L}{2\pi v}}%
~e^{-\frac{L\left( \bar{n}-n_{av}\right) ^{2}}{2v}}
\end{equation}%
Note that we have included an explicit list of the parameters which control
the distribution ${\cal \tilde{P}}$.

\subsection{Statistics of variances}

Next, we turn to the central issue of this paper. If $p$ is estimated
through the variance of a single string of $L$ observations ($n_{i}$), how
often can we expect the estimate to be complex? In other words, what is the
probability that the variance$\,L^{-1}\sum_{i}n_{i}^{2}-\bar{n}^{2}$ exceeds
the absolute bound 
\begin{equation}
V_{abs}\equiv N/4\,\,?  \label{Vbound}
\end{equation}%
To answer this question, we focus on the probability that $%
L^{-1}\sum_{i}n_{i}^{2}-\bar{n}^{2}$ assumes a given value, say, $V$. First,
we ask for the (joint) probability that the number of right steps, observed
over $L$ nights, takes the values $n_{1}$, $n_{2}$,..., $n_{L}$. Since the
events of any night are independent of all other nights, this joint
probability $P\left( n_{1},n_{2},...,n_{L}\right) $ $\,$is just the product
of the probabilities for a single night, i.e., $\prod_{\ell =1}^{L}P(n_{\ell
})$ -- and $P(n_{\ell })$ is simply given by the binomial, Eqn (\ref{Pn}).
Summing over all possible outcomes $\left\{ n_{\ell }\right\} \equiv \left\{
n_{1},n_{2},...,n_{L}\right\} $ by using a Kronecker delta to count only
those for which the variance equals $V$, we obtain 
\begin{equation}
{\cal P}\left( V\right) =\sum_{\left\{ n_{_{\ell }}\right\} }\delta \left( V,%
\frac{1}{L}\sum_{i=1}^{L}n_{i}^{2}-\left[ \frac{1}{L}\sum_{i=1}^{L}n_{i}%
\right] ^{2}\right) \prod_{\ell =1}^{L}P\left( n_{\ell }\right) \,\,.
\label{P-of-V}
\end{equation}%
Unfortunately, we are unable to evaluate this expression exactly. However,
an excellent approximation can be obtained if $N$ is not too small (as in
the cases posed in Ref.~4) so that we can use the Gaussian
approximation, Eqn (\ref{PGn}), instead of the exact binomial, Eqn (\ref{Pn}%
). A reader intimately familiar with error analysis will recognize our quest
as the probability density function for the $\chi ^{2}$-distribution \cite%
{chi-sq+gamma}. For pedagogical purposes, let us show how to make progress
from this point. Let the probability that the variance lies in the interval $%
\left[ V,V+dV\right] $ be ${\cal \tilde{P}}\left( V\right) dV$. Then, 
\begin{eqnarray}
{\cal \tilde{P}}\left( V;v,L\right)  &=&\prod_{\ell =1}^{L}\int_{-\infty
}^{\infty }dn_{\ell }\tilde{P}\left( n_{\ell }\right) \times   \nonumber \\
&\delta &\left( V-\frac{1}{L}\sum_{\ell =1}^{L}n_{\ell }^{2}+\left[ \frac{1}{%
L}\sum_{\ell =1}^{L}n_{\ell }\right] ^{2}\right) \,\,,  \label{Ptilde}
\end{eqnarray}%
where we have included the relevant parameters ($v,L$) explicitly. Note that
all variables are now continuous with infinite range and, correspondingly,
the $\delta $ here is the Dirac delta. To evaluate this distribution,
consider its Laplace transform 
\begin{eqnarray}
{\cal L}\left( \mu \right)  &\equiv &\int dVe^{-\mu V}{\cal \tilde{P}}\left(
V;v,L\right)   \nonumber \\
&=&\prod_{\ell =1}^{L}\int_{-\infty }^{\infty }dn_{\ell }\tilde{P}\left(
n_{\ell }\right) \times   \nonumber \\
&\exp &\left[ -\frac{\mu }{L}\sum_{\ell =1}^{L}n_{\ell }^{2}+\frac{\mu }{%
L^{2}}\left( \sum_{\ell =1}^{L}n_{\ell }\right) ^{2}\right] \,\,.
\label{L-mu}
\end{eqnarray}%
Since each $\tilde{P}$ is a Gaussian, we have here a generalized Gaussian
integral. Explicitly, we have 
\begin{eqnarray}
&&{\cal L}\left( \mu \right) =\prod_{\ell =1}^{L}\int_{-\infty }^{\infty }%
\frac{dn_{\ell }}{\sqrt{2\pi v}}\times  \\
&\exp &\left[ -\frac{1}{2v}\sum_{\ell =1}^{L}\left( n_{\ell }-Np\right) ^{2}-%
\frac{\mu }{L}\sum_{\ell =1}^{L}n_{\ell }^{2}+\frac{\mu }{L^{2}}\left(
\sum_{\ell =1}^{L}n_{\ell }\right) ^{2}\right]   \nonumber
\end{eqnarray}%
To proceed, it is best to displace each $n_{\ell }$ by $Np$. Verifying that
the variance is independent of such a shift, we have a simpler form: 
\begin{eqnarray}
{\cal L}\left( \mu \right)  &=&\prod_{\ell =1}^{L}\int_{-\infty }^{\infty }%
\frac{dn_{\ell }}{\sqrt{2\pi v}}\times  \\
&\exp &\left[ -\left( \frac{1}{2v}+\frac{\mu }{L}\right) \sum_{\ell
=1}^{L}n_{\ell }^{2}+\frac{\mu }{L^{2}}\left( \sum_{\ell =1}^{L}n_{\ell
}\right) ^{2}\right]   \nonumber
\end{eqnarray}%
Rescaling each $n_{\ell }$ by $\sqrt{vL/\left( L+2\mu v\right) }$ simplifies
this expression further: 
\begin{eqnarray}
{\cal L}\left( \mu \right)  &=&\left( \frac{L}{L+2\mu v}\right)
^{L/2}\prod_{\ell =1}^{L}\int_{-\infty }^{\infty }\frac{dn_{\ell }}{\sqrt{%
2\pi }}\times   \label{L} \\
&\exp &\left\{ -\frac{1}{2}\left[ \sum_{\ell =1}^{L}n_{\ell }^{2}-\frac{2\mu
v}{\left( L+2\mu v\right) }\left( \frac{1}{\sqrt{L}}\sum_{\ell
=1}^{L}n_{\ell }\right) ^{2}\right] \right\}   \nonumber
\end{eqnarray}%
The exponent should be regarded as a quadratic form 
\begin{equation}
-\frac{1}{2}\sum n_{\ell }Q_{\ell \ell ^{\prime }}n_{\ell ^{\prime }}
\end{equation}%
where the matrix is just an identity matrix plus a term proportional to the
tensor product of a single unit vector, namely, $\left( 1,1,...,1\right) /%
\sqrt{L}$. The eigenvalues of such a matrix are all unity, except for one,
which is unity plus the value of this proportionality constant. In Eqn (\ref%
{L}), we have purposefully written the last term in the exponent to display
this constant, i.e., $-2\mu v/\left( L+2\mu v\right) $. Exploiting 
\begin{equation}
\prod_{\ell =1}^{L}\int_{-\infty }^{\infty }\frac{dn_{\ell }}{\sqrt{2\pi }}%
\exp \left[ -\frac{1}{2}\sum n_{\ell }Q_{\ell \ell ^{\prime }}n_{\ell
^{\prime }}\right] =\left( \det Q\right) ^{-1/2}\,\,,
\end{equation}%
we have 
\begin{eqnarray}
{\cal L}\left( \mu \right)  &=&\left( \frac{L}{L+2\mu v}\right) ^{L/2}\left(
1-\frac{2\mu v}{\left( L+2\mu v\right) }\right) ^{-1/2}  \nonumber \\
&=&\left( \frac{L}{L+2\mu v}\right) ^{\left( L-1\right) /2}.  \label{L-ans}
\end{eqnarray}%
To obtain ${\cal \tilde{P}}\left( V;v,L\right) $, we only need to perform an
inverse Laplace transform. Fortunately, ${\cal L}\left( \mu \right) $ is
simple enough to appear in standard tables \cite{AS} and the explicit answer
is 
\begin{equation}
{\cal \tilde{P}}\left( V;v,L\right) =\frac{1}{V\Gamma \left( \frac{L-1}{2}%
\right) }\left( \frac{LV}{2v}\right) ^{\left( L-1\right) /2}e^{-LV/2v}\,\,.
\label{Ans}
\end{equation}%
Of course, this distribution may be written in scaling form: 
\begin{equation}
{\cal \tilde{P}}\left( V;v,L\right) =\frac{L}{2v}\Phi _{\frac{L-1}{2}}\left(
x\right)   \label{scale}
\end{equation}%
where the (non-negative) scaling variable is 
\begin{equation}
x\equiv \frac{LV}{2v}\,\,,  \label{x}
\end{equation}%
and $\Phi $ is the standard gamma distribution \cite{chi-sq+gamma}: 
\begin{equation}
\Phi _{\gamma }\left( x\right) =\frac{x^{\gamma -1}}{\Gamma (\gamma )}e^{-x}.
\label{Phi}
\end{equation}%
with 
\[
\gamma \equiv (L-1)/2
\]%
for our case. We briefly note some of its properties: It peaks at $%
x_{peak}=\max \left\{ 0,\gamma -1\right\} $ and has moments $\left\langle
x^{n}\right\rangle =$ $\Gamma (n+\gamma )/\Gamma (\gamma )$. Thus, its
average and variance both equal $\gamma $: $\left\langle x\right\rangle
=\left\langle x^{2}\right\rangle -\left\langle x\right\rangle ^{2}=$ $\gamma 
$.

Returning to the problem we posed for the human behaviorist, namely, to
compile a histogram for the set of $V_{\alpha }$, Eqn (\ref{nV}), on the
basis of data for $M$ drunkards, Eqn (\ref{nai}), we should find 
\begin{equation}
H\left( V\right) \rightarrow M{\cal \tilde{P}}\left( V;v,L\right) 
\end{equation}%
for very large $M$.

Before applying this result to the problem quoted above, let us comment on
several interesting features.

From Eqn (\ref{x}), we might infer that the relevant scale for $V$ is $%
v/L=Np\left( 1-p\right) /L$. Nonetheless, the {\em average }(or expectation
value) of $V$ , being 
\begin{equation}
V_{av}=\frac{2v}{L}\gamma =Np\left( 1-p\right) \left[ 1-\frac{1}{L}\right]
\,\,,  \label{Vav}
\end{equation}%
is much closer to $v$, especially for large $L$. Another interesting
quantity is the {\em most likely} value for $V$, which is simply related to
the peak value of $\Phi _{\gamma }$, via $V_{peak}=$ $\left( 2v/L\right)
x_{peak}$. So, 
\begin{equation}
V_{peak}=\left\{ 
\begin{tabular}{ll}
$0$ & for $L\leq 3$ \\ 
$Np\left( 1-p\right) \left[ 1-\frac{3}{L}\right] $ & for $L>3$%
\end{tabular}%
\ \ \right. \,\,.  \label{Vpk}
\end{equation}%
Both are consistently less than $Np\left( 1-p\right) $, the ``true
variance''. Though both approach $v$ monotonically as $L$ increases, these
results provide a measure of how much the variance of a short data string
(i.e., small $L$) can differ from its asymptotic ($L\rightarrow \infty $)
value.

Since short data strings lead to the most serious discrepancies, let us make
a detour for the cases $L=1,2,$ and $3$. The appearance of $\Gamma \left( 
\frac{L-1}{2}\right) $ in the denominator of Eqn (\ref{Ans}) ensures that,
for $L=1$, ${\cal \tilde{P}}\equiv 0$ for $V\neq 0$. On the other hand, this
distribution is normalized, so that we may conclude ${\cal \tilde{P}}\left(
V;v,0\right) =\delta \left( V\right) $, regardless of $v$. This result is
perfectly understandable, since $L=1$ corresponds to our observing the
drunkard for only a {\em single} night. If a ``data set'' consists of only a 
{\em single} number, the ``variance'' is necessarily zero, regardless of the
behavior of the drunkard! For $L=2$, ${\cal \tilde{P}}$ diverges at the
origin. There is no cause for alarm, however, as the divergence is weak
enough for $\int {\cal \tilde{P}}dV$ to be finite in any neighborhood of $V=0
$. Indeed, a better perspective is provided by the distribution for the
standard deviation 
\[
\sigma \equiv \sqrt{V}\,
\]%
Then, we find a familiar looking expression:  
\[
P\left( \sigma ;v,2\right) \equiv \frac{dV}{d\sigma }{\cal \tilde{P}}\left(
V;v,2\right) =\frac{2}{\sqrt{\pi v}}e^{-\sigma ^{2}/v}.
\]%
However, keep in mind that we have only ``half a Gaussian'' here, since $%
\sigma \in \left[ 0,\infty \right] $! Finally, the other curious case is $L=3
$ where ${\cal P}\left( V;v,3\right) $ is a pure exponential, for which the
standard deviation assumes the same value as the average. By contrast, the
situation for large $L$ provides few surprises. The gamma distribution
approaches a Gaussian with width $O\left( 1/\sqrt{L}\right) $.

\section{Summary and application to the Gould-Tobochnik\protect\cite{GT}
problem}

Let us summarize our findings before applying them to the problem quoted in
the Introduction. For the reader's convenience, we reiterate the set-up:
Consider an ``ensemble'' of $M$ {\em identical} drunkards with $p$ as the
probability of stepping to the right. Each is observed for $L$ nights,
taking $N$ total steps each night. For drunkard $\alpha $ ($=1,...,M$), the
number of right steps is recorded for each night ($i=1,...L$) and denoted by 
$n_{\alpha ,i}\,.$ From these, we compute the averages $\bar{n}_{\alpha
}\equiv L^{-1}\sum_{i}n_{\alpha ,i}$ and variances $V_{\alpha }\equiv
L^{-1}\sum_{i}n_{\alpha ,i}^{2}-\left( \bar{n}_{\alpha }\right) ^{2}$.
Normalized histograms for these are compiled and denoted by $H\left( \bar{n}%
\right) /M$ and $H\left( V\right) /M$. We find that, already for moderate $N$ (such
as 20), the first will approach a Gaussian distribution while the second, a
gamma distribution. Explicitly, 
\begin{equation}
\frac{H\left( \bar{n}\right) }{M}%
\mathrel{\mathop{\rightarrow }\limits_{M\rightarrow \infty }}%
{\cal \tilde{P}}\left( \bar{n};n_{av},v,L\right) =\sqrt{\frac{L}{2\pi v}}%
~e^{-\frac{L\left( \bar{n}-n_{av}\right) ^{2}}{2v}}
\end{equation}%
and 
\begin{equation}
\frac{H\left( V\right) }{M}%
\mathrel{\mathop{\rightarrow }\limits_{M\rightarrow \infty }}%
{\cal \tilde{P}}\left( V;v,L\right) =\frac{x^{\gamma }e^{-x}}{V\Gamma \left(
\gamma \right) }\,\,.
\end{equation}%
where $n_{av}\equiv Np$, $v\equiv Np\left( 1-p\right) $, $x\equiv LV/2v$,
and $\gamma \equiv \left( L-1\right) /2.$

Applying these results to the specific problem at hand, let us {\em assume}
that the underlying $p$ is $0.6$. With $L=10$ and $N=20$ or $100$, we have
the full distribution of variances 
\begin{equation}
{\cal \tilde{P}}\left( V;0.24N,10\right) =\frac{x^{4.5}e^{-x}}{V\Gamma
\left( 4.5\right) }.  \label{P10}
\end{equation}
with 
\begin{equation}
x=\frac V{0.048N}.
\end{equation}
As noted in the previous section, the peak of ${\cal \tilde{P}}$ occurs at $v%
\left[ 1-3/L\right] $, which is 
\begin{equation}
V_{peak}\cong 0.168N
\end{equation}
in this case. Although this value appears far below the maximum allowed, $%
V_{abs}=0.25N$, we should be more careful with our original question:

\begin{center}
{\bf How likely will $V$ exceed $V_{abs}$?}
\end{center}

To answer this question, we consider the integral 
\begin{equation}
\rho \left( p,L\right) \equiv prob\left( V>V_{abs}\right)
=\int_{N/4}^{\infty }{\cal \tilde{P}}\left( V;v,L\right) dV\,\,,
\label{prob}
\end{equation}%
Using the scaled variable, $\rho $ is just 
\begin{equation}
\rho \left( p,L\right) =\frac{1}{\Gamma (\gamma )}\int_{x_{abs}}^{\infty
}x^{\gamma -1}e^{-x}dx  \label{rho}
\end{equation}%
which should be recognized as a standard $\chi ^{2}$-probability function %
\cite{chi-sq+gamma} evaluated at a particular point: 
\[
x_{abs}=\frac{LV_{abs}}{2Np\left( 1-p\right) }=\frac{\gamma +1/2}{4p\left(
1-p\right) }\,\,.
\]%
Note that $\rho $ is now manifestly {\em independent} of $N$, the number of
steps taken by each drunk per night. As a result, we have written explicitly
in Eqn (\ref{rho}) that it depends solely on $p$ and $L$.

Applying this integral to the case at hand, we write 
\begin{equation}
\rho \left( 0.6,10\right) \cong \frac{1}{\Gamma \left( 4.5\right) }%
\int_{5.2083}^{\infty }x^{3.5}e^{-x}dx  \label{rho-here}
\end{equation}%
and, computing the integral numerically, find 
\begin{equation}
\rho \left( 0.6,10\right) \cong 0.3178\,\,.  \label{3178}
\end{equation}%
This is an astonishingly large value! In other words, by observing $M$
drunkards (identical ones, with 0.6 probability of stepping to the right)
for $10$ nights each, about $M/3$ of the data sets will lead to a variance
exceeding the absolute bound of $N/4$! In the equivalent scenario, if the
string of numbers in the text \cite{GT} were generated anew with each access
to the URL, then about {\em a third} of the students would find complex $%
p_{\sigma }$'s by using $\sigma _{n}$, Eqn (\ref{sn}). Let us remind the
reader that this conclusion is {\em independent }of $N$!

For students to appreciate these considerations better, it suffices to set 
up a crude Excel spreadsheet. Assuming $p=0.6$ and placing the formula 
IF(RAND()<0.6,1,0) in, e.g., the block A1-T12000, 
we can ``create'' the data (with 
right/left steps recorded as 1/0) for $M=1200$ drunkards, observed on $L=10$ 
nights (10 successive rows for each drunkard), for $N=20$ steps (columns 
A-T). Summing each row into column U, we have the ``data set'' 
$\left\{ n_{\alpha ,i}\,\right\} $. 
After computing the averages and variances for 
the numbers in column U (in blocks of 10), the histogram function can be 
invoked to provide $H\left( \bar{n}\right) $ and $H\left( V\right) $. 
In Fig.~1, we show the result for the latter (for a particular run, of course). 
Since the peak position is expected to be at $3.36$ (considerably less than 
the ``true variance'' of $4.8$!), we choose bins of width 0.4, 
so that the center of one bin (3.4) approximately coincides with the peak. 
As we see from the histogram, the peak frequency indeed lies in the appropriate 
bin. 
In Fig.~1, we also plot the theoretical 
distribution, Eqn (\ref{P10}), multiplied by $M=1200$. The zero-parameter 
``fit'' is clearly excellent. Since the peak is quite far from the average, 
we have also computed the latter and find $V_{av}\cong 4.331$. This value is 
entirely consistent with the predicted 
$\left( 0.24\right) \left( 20\right) \left( 0.9\right) =4.32$, 
its 
approximate location being indicated by an 
arrow in Fig.~1. Lastly, we program the spread sheet to find the number
of $V$'s lying above the bound of $5$ in this sample. The result, for this 
particular run was $374$, again consistent with the predicted $381=(0.3178)(1200).$

\begin{figure}[tbp]
\input{epsf}
\begin{center}
\vspace{-3.5cm}
\begin{minipage}{0.7\textwidth}
  \epsfxsize = 0.7\textwidth \epsfysize = 0.7\textwidth 
  \epsfbox{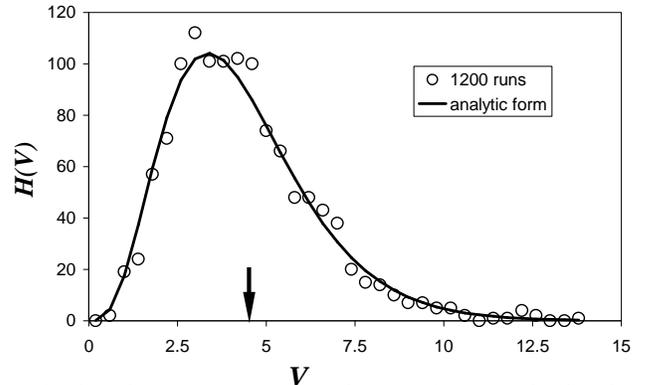}
    \vspace{-0.5cm}
\end{minipage}
\end{center}
\caption{Histogram and analytic form for the distribution of variances. The
average variance, $V_{av}\simeq 4.3$, is indicated by the arrow. The data
are ``observations'' of 1200 drunkards for 10 nights, each taking 20 steps
each night.}
\end{figure}

Given this large percentage of obviously non-sensical estimates, a
conscientious observer would naturally attempt to improve the situation, by
making more observations. Noting that this difficulty is {\em independent}
of $N$ (the number of steps tallied on each night), our observer might try
increasing $L$, by watching the drunks for more than 10 nights. One might
hope that taking data for $100$ nights, say, should surely lead to a much
larger proportion of ``good data'' (i.e., variances that lead to real
estimates for $p$). Mustering extreme patience, our observer collects the
necessary data, performs the analysis and is absolutely shocked to discover
the result: instead of dropping sharply, $\rho $ has {\em increased even
further}, to $0.3416$! In fact, a little thought shows that $\rho $ cannot
be a monotonically decreasing function of $L$. Indeed, for $L=1$, $\rho
\left( p,1\right) $ {\em vanishes}, since ${\cal \tilde{P}}=\delta \left(
V\right) $. So, at least initially, $\rho \left( 0.6,L\right) $ increases
with $L$. In fact, it peaks at $L=41$ where it takes its maximum value of $%
0.35553$. Beyond $L=41$, it decreases, but so slowly that $\rho \left(
0.6,10^{2}\right) \cong 0.3416$ is still larger than $\rho \left(
0.6,10\right) \cong 0.3178$. Indeed, it decreases exceedingly slowly, so
that, e.g., $\rho \left( 0.6,10^{3}\right) \cong 17\%$!

\begin{figure}[tbp]
\input{epsf}
\begin{center}
\vspace{-3.5cm}
\begin{minipage}{0.7\textwidth}
  \epsfxsize = 0.68\textwidth \epsfysize = 0.7\textwidth 
  \epsfbox{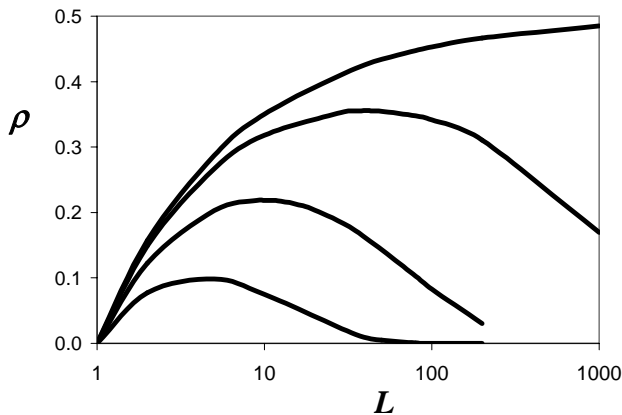}
    \vspace{-0.5cm}
\end{minipage}
\end{center}
\caption{The probability $\protect\rho \left( p,L\right)$ that the variance
will exceed the absolute bound $N/4$, plotted vs $L$ for various $p$. The
curves, from the top, are for $p=0.5,0.6,0.7,$ and $0.8$, respectively.}
\end{figure}

Since $\rho $ is identically zero at $L=1$ (for any $p$), while $\rho \left(
p,L\rightarrow \infty \right) $ is expected to vanish, an interesting
question is the following. Given $p$, for what value of $L$ will $\rho $
reach its maximum? Though easy to pose, we have not found a simple analytic
answer to this question. Instead, in Fig.~2, we present plots of $\rho
\left( p,L\right) $ for a few $p$'s, so that our readers can appreciate its
general behavior. Let us end by simply stating another remarkable result:
for the {\em unbiased }walk, $\rho \left( 0.5,L\right) $ is a {\em %
monotonically increasing} function of $L$, with limiting value of $1/2$. In
other words, as the number of observation nights is increased, the
likelyhood of the computed variance exceeding the true one {\em increases}.
In the limit of compiling data from ``infinitely many'' nights (of $N$-step
walks), it is equally likely for the variance to exceed $N/4$ (the true
value) as otherwise! For students interested in critical phenomena, it is
possible to regard $\left( p,L\right) =\left( 0.5,\infty \right) $ as a
critical point and to seek a scaling function for $\rho $ involving $\left(
p-0.5\right) L^{1/4}$. A full analysis of these issues would be more
suitable, we believe, for the readership of another journal.

\section{Concluding remarks}

We have analyzed an interesting problem posed in the manuscript of Gould and
Tobochnik \cite{GT}. A random walker is observed for $10$ nights, taking $20$
steps each night. The number of {\em right} steps taken each night is
recorded, and the student is asked to compute the average and variance from,
and for, these 10 data points. Letting $p$ denote the probability of the
walker taking a right step, the true variance is just $20p(1-p)$ and can
never exceed $5$. Nevertheless, this bound {\em is exceeded} for the
specific sequence of ten numbers in the text! In an effort to resolve this
surprisingly result, we find that, if many such sequences (of ten numbers)
were generated according to $p=0.6$, nearly 32\% of them will lead to a
variance exceeding the absolute bound. Indeed, we found the full
distribution for the variance, ${\cal \tilde{P}}\left( V;L,N,p\right) $, for
the general case ($L$ nights, $N$ steps, arbitrary $p$).

There is an alternative perspective to the behavior discussed here, namely,
in terms of the statistics associated with ``binning.'' After all, if each
step of the random walker is independent, then -- instead of binning the
data into $L$ nights of $N$ steps each -- we might as well consider the
whole data set as a single string of $NL$ steps. Associated with this string
is, of course, just a single number: $\tilde{n}$, the total number of right
steps. Seeking, as before, the ``average'' and ``variance'' of this long
string, the ``average'' is simply $\tilde{n}$, giving just {\em one}
estimate for $p$, i.e., $\tilde{n}/NL$. Moreover, the ``variance'' is
necessarily zero. Now, by observing $M$ drunkards, we can make a histogram
of the associated estimates for $p$. For large $M$, this histogram will
approach the theoretical distribution of a binomial associated with $%
{NL \choose \tilde{n}}%
$. If $NL$ is also large, the histogram is well approximated by a Gaussian
centered on $p$ with standard deviation $\sqrt{p\left( 1-p\right) /NL}$. To
write this string in the setting of the textbook problem, we would {\em bin}
the $NL$ steps into $L$ bins and arrive at $L$ numbers $n_i$, as well as $L$
estimates (for $p$): $n_i/L$. The average of these estimates is {\em exactly 
}$\tilde{n}/NL$, i.e., as if no binning took place. However, by binning, we
have created a non-trivial ``variance,'' associated with the binned values $%
n_i$! Since it will depend on the bin size (or equivalently, the number of
bins: $L$), we should denote such a variance by $V_L$. Above, we have
presented its distribution and its non-trivial dependence on $L$.

Our goal in this article is to demonstrate that the full {\em distribution}
of a physical quantity often carries extremely interesting information and
can be invoked to put surprising properties of individual moments into
perspective. We believe that this message is central to statistical
mechanics and should be taught within the context of upper level courses. Of
course, the notion of full distributions {\em is} frequently included in
texts but typically limited to the binomial and Gaussian distributions. In
the context of the random walk and related problems, these lead naturally to
a discussion of the central limit theorem. This theorem elevates the
Gaussian up to the status of a ``universal'' distribution. The appeal of
this notion is so strong that students might be lulled into thinking that 
{\em all} distributions are ``normal'' or ``bell shaped,'' for which only
average and variance are needed. Yet, many distributions are not normal and
display a variety of interesting properties. A good example is the Poisson
distribution, included in GT \cite{GT} to convince readers that they need
not be afraid of flying. For these reasons, we believe that it would be
valuable to devote some lessons in a course on statistical mechanics, or an
entire section of a graduate or an undergraduate text, to distributions. In
particular, the notion of derived, or induced distributions is very
important: In many areas of physics, we are frequently interested in
stochastic quantities which are themselves {\em specific functions }of
other, underlying random variables (often assumed to be Gaussian). Of
course, finding the full distribution of such quantities can turn into a
very challenging task but the rewards can be considerable, as we seek to
demonstrate with two examples, both of which are relatives of the simple
one-dimensional random walk. The first, concerning the statistical widths of
a one-dimensional interface leads to a {\em universal} (but non-Gaussian)
distribution, describing a {\em whole class} of interface models. The second
illustrates the observation that, depending on the question posed, even
simple random walks can give rise to non-analyticities which may be
difficult to explore fully.

{\em Statistical widths of a one-dimensional\ interface, or average
deviations in a random walk.} The simple random walk on a line is just a
string of R and L steps. Plotting the displacement on the $y$-axis and the
number $i$ of steps taken on the abscissa, a particular walk can be viewed
as a specific configuration of a {\em one-dimensional} interface, embedded
in a two-dimensional bulk: the interface {\em height} $h_{i}$ is measured
along $y$ vs the ``column label'' $i=1,2,...N$. In the language of
interfaces, two natural and frequently considered quantities are the average
height $\bar{h}\equiv \sum_{i}h_{i}/N$ and width (easily recognized as a
standard deviation) $w\equiv \sqrt{\sum_{i}\left( h_{i}-\bar{h}\right) ^{2}/N%
}$ of a given interface. However, in the context of random walks, neither of
these quantities comes to mind easily. Instead of being an ensemble average
over very many walks, $\bar{h}$ is the average displacement over {\em time},
of a {\em specific} walk in a {\em specific} night, and hence depends on the
entire history of this particular walk. Similarly, $w$ is different from one
interface to another and, in terms of the drunkard, typically changes from
one night to the next. Here, we are interested in the full distributions of $%
\bar{h}$ and $w$. Obviously, we can compile their histograms, $H\left( \bar{h%
}\right) $ and $H\left( w\right) $, by generating many interface
configurations or by observing the drunkard over many nights. If the
underlying probability is symmetric ($p\left( \pm 1\right) =1/2$), then it
is easy to imagine that $H\left( \bar{h}\right) $ is symmetric and
approaches a Gaussian, thanks to the central limit theorem. On the other
hand, it is not so easy to guess what $H\left( w\right) $ should be. First, $%
w$ is never negative, so $H\left( w\right) $ must ``end'' at $w=0$. Second,
though the average (and perhaps the peak position) should increase with $%
\sqrt{N}$, the ``top end'' of $H\left( w\right) $ will be $O\left( N\right) $%
. Referring to Ref.~8 for details, we just summarize the key results.
Noting that $\left\langle w^{2}\right\rangle $, the width-square {\em %
averaged} over all possible configurations, is just $L/12$, we obtain $%
H\left( w\right) \sim \exp \left[ -\pi ^{2}w^{2}/\left\langle
w^{2}\right\rangle \right] $ for large widths and $H\left( w\right) \sim
w^{-5}\exp \left[ -3\left\langle w^{2}\right\rangle /2w^{2}\right] $ for
small $w$'s. In other words, configurations with widths significantly
larger/smaller than $\left\langle w^{2}\right\rangle $ are exponentially
suppressed. Remarkably, this distribution is {\em universal}, describing a
whole class of interface models, in the same sense that the Gaussian is a
limiting distribution for sums of random numbers.

{\em Distribution of longest returns. }Here, we turn to an extremely
interesting example, in which there is no analytic solution even for simple
one-dimensional random walks. Motivated by the physics of charged polymers
(a string of monomers, each carrying charge $\pm 1$), Ertas and Kantor \cite%
{EK} explored the properties of the largest neutral segments. Translated
into the language of one-dimensional random walks, a neutral segment,
consisting of equal numbers of opposite charges, corresponds to a part of
the walk where the walker returns to a particular site. So, the ``largest
neutral segment'' maps into the ``longest return path,'' or largest number
of steps between visits to the same site (regardless of which site). A more
explicit phrasing of the question is: ``Of all the $2^{N}$ random walks of $N
$ steps on a line, how many have $N^{\prime }$ (with $0\leq N^{\prime }\leq N
$) as the longest return path?'' Denoting the answer by $H\left( N^{\prime
};N\right) $, the normalized distribution is $P\left( N^{\prime }\right)
=H\left( N^{\prime };N\right) /2^{N}$. Again, it may be helpful to consider
the extremes. For $N^{\prime }=0$, there are precisely two walks: all right
or all left steps. So, $H\left( 0;N\right) =2$. At the other extreme, for $%
N^{\prime }=N$, these walks are the familiar ones which return to the
starting point, so that $H\left( N;N\right) =%
{N \choose N/2}%
$. In the limit of large $N$, it is better to use the fraction $\phi \equiv
N^{\prime }/N$ as a variable and to take the continuum limit, so that an
appropriate probability density, $p\left( \phi \right) $, emerges from $%
P\left( N^{\prime }\right) $. Remarkably, $p\left( \phi \right) $ develops a
kink, i.e., a {\em discontinuous} first derivative, at $\phi =1/2$ (Fig.~3
in Ref.~9)! Despite the simple sounding nature of this question, it is
clearly quite a complex issue. Though some understanding of this unexpected
phenomenon is possible, it is far from being completely solved.

To conclude, we hope to have motivated teachers and students alike to think
beyond the first few moments and to focus on full distributions, if at all
possible. To whet our readers' appetites, we have presented
a particularly striking example in the context of simple random walks: if
one uses finite data strings to estimate the (asymptotic) probability $p$,
the estimate can easily turn out to be manifestly nonsensical, namely,
complex! Beyond this ``demonstration'', we briefly discussed two further
examples, all associated with random walks. Hopefully, these concepts will
challenge our readers to explore, or discover, their own favorite
distributions.

\vspace{0.5cm}

{\bf Acknowledgements.} We thank Harvey Gould, Jan Tobochnik, and Uwe T\"{a}%
uber for very helpful discussions. This work is supported in part by a grant
from the National Science Foundation, through the Division of Materials
Research.

\end{document}

%% file: epsf.tex
\ifx\epsfannounce\undefined \def\epsfannounce{\immediate\write16}\fi
 \epsfannounce{This is `epsf.tex' v2.7k <10 July 1997>}%
\newread\epsffilein    
\newif\ifepsfatend     
\newif\ifepsfbbfound   
\newif\ifepsfdraft     
\newif\ifepsffileok    
\newif\ifepsfframe     
\newif\ifepsfshow      
\epsfshowtrue          
\newif\ifepsfshowfilename 
\newif\ifepsfverbose   
\newdimen\epsfframemargin 
\newdimen\epsfframethickness 
\newdimen\epsfrsize    
\newdimen\epsftmp      
\newdimen\epsftsize    
\newdimen\epsfxsize    
\newdimen\epsfysize    
\newdimen\pspoints     
\pspoints = 1bp        
\epsfxsize = 0pt       
\epsfysize = 0pt       
\epsfframemargin = 0pt 
\epsfframethickness = 0.4pt 
\def\epsfbox#1{\global\def\epsfllx{72}\global\def\epsflly{72}%
   \global\def\epsfurx{540}\global\def\epsfury{720}%
   \def\lbracket{[}\def\testit{#1}\ifx\testit\lbracket
   \let\next=\epsfgetlitbb\else\let\next=\epsfnormal\fi\next{#1}}%
%
%
\def\epsfgetlitbb#1#2 #3 #4 #5]#6{%
   \epsfgrab #2 #3 #4 #5 .\\%
   \epsfsetsize
   \epsfstatus{#6}%
   \epsfsetgraph{#6}%
}%
\def\epsfnormal#1{%
    \epsfgetbb{#1}%
    \epsfsetgraph{#1}%
}%
\newhelp\epsfnoopenhelp{The PostScript image file must be findable by
TeX, i.e., somewhere in the TEXINPUTS (or equivalent) path.}%
\def\epsfgetbb#1{%
%
%
    \openin\epsffilein=#1
    \ifeof\epsffilein
        \errhelp = \epsfnoopenhelp
        \errmessage{Could not open file #1, ignoring it}%
    \else                       
        {
            \chardef\other=12
            \def\do##1{\catcode`##1=\other}%
            \dospecials
            \catcode`\ =10
            \epsffileoktrue         
            \epsfatendfalse     
            \loop               
                \read\epsffilein to \epsffileline
                \ifeof\epsffilein 
                \epsffileokfalse 
            \else                
                \expandafter\epsfaux\epsffileline:. \\%
            \fi
            \ifepsffileok
            \repeat
            \ifepsfbbfound
            \else
                \ifepsfverbose
                    \immediate\write16{No BoundingBox comment found in %
                                    file #1; using defaults}%
                \fi
            \fi
        }
        \closein\epsffilein
    \fi                         
    \epsfsetsize                
    \epsfstatus{#1}%
}%
%
\def\epsfclipon{\def\epsfclipstring{ clip}}%
\def\epsfclipoff{\def\epsfclipstring{\ifepsfdraft\space clip\fi}}%
\epsfclipoff 
%
%
\def\epsfspecial#1{%
     \epsftmp=10\epsfxsize
     \divide\epsftmp\pspoints
     \ifnum\epsfrsize=0\relax
       \special{PSfile=\ifepsfdraft psdraft.ps\else#1\fi\space
                llx=\epsfllx\space
                lly=\epsflly\space
                urx=\epsfurx\space
                ury=\epsfury\space
                rwi=\number\epsftmp
                \epsfclipstring
               }%
     \else
       \epsfrsize=10\epsfysize
       \divide\epsfrsize\pspoints
       \special{PSfile=\ifepsfdraft psdraft.ps\else#1\fi\space
                llx=\epsfllx\space
                lly=\epsflly\space
                urx=\epsfurx\space
                ury=\epsfury\space
                rwi=\number\epsftmp
                rhi=\number\epsfrsize
                \epsfclipstring
               }%
     \fi
}%
%
\def\epsfframe#1%
{%
  \leavevmode                   
  \setbox0 = \hbox{#1}%
  \dimen0 = \wd0                                
  \advance \dimen0 by 2\epsfframemargin         
  \advance \dimen0 by 2\epsfframethickness      
  \vbox
  {%
    \hrule height \epsfframethickness depth 0pt
    \hbox to \dimen0
    {%
      \hss
      \vrule width \epsfframethickness
      \kern \epsfframemargin
      \vbox {\kern \epsfframemargin \box0 \kern \epsfframemargin }%
      \kern \epsfframemargin
      \vrule width \epsfframethickness
      \hss
    }
    \hrule height 0pt depth \epsfframethickness
  }
}%
\def\epsfsetgraph#1%
{%
   %
   %
   \leavevmode
   \hbox{
     \ifepsfframe\expandafter\epsfframe\fi
     {\vbox to\epsfysize
     {%
        \ifepsfshow
            \vfil
            \hbox to \epsfxsize{\epsfspecial{#1}\hfil}%
        \else
            \vfil
            \hbox to\epsfxsize{%
               \hss
               \ifepsfshowfilename
               {%
                  \epsfframemargin=3pt 
                  \epsfframe{{\tt #1}}%
               }%
               \fi
               \hss
            }%
            \vfil
        \fi
     }%
   }}%
   %
   %
   \global\epsfxsize=0pt
   \global\epsfysize=0pt
}%
%
%
\def\epsfsetsize
{%
   \epsfrsize=\epsfury\pspoints
   \advance\epsfrsize by-\epsflly\pspoints
   \epsftsize=\epsfurx\pspoints
   \advance\epsftsize by-\epsfllx\pspoints
%
%
   \epsfxsize=\epsfsize{\epsftsize}{\epsfrsize}%
   \ifnum \epsfxsize=0
      \ifnum \epsfysize=0
        \epsfxsize=\epsftsize
        \epsfysize=\epsfrsize
        \epsfrsize=0pt
%
%
      \else
        \epsftmp=\epsftsize \divide\epsftmp\epsfrsize
        \epsfxsize=\epsfysize \multiply\epsfxsize\epsftmp
        \multiply\epsftmp\epsfrsize \advance\epsftsize-\epsftmp
        \epsftmp=\epsfysize
        \loop \advance\epsftsize\epsftsize \divide\epsftmp 2
        \ifnum \epsftmp>0
           \ifnum \epsftsize<\epsfrsize
           \else
              \advance\epsftsize-\epsfrsize \advance\epsfxsize\epsftmp
           \fi
        \repeat
        \epsfrsize=0pt
      \fi
   \else
     \ifnum \epsfysize=0
       \epsftmp=\epsfrsize \divide\epsftmp\epsftsize
       \epsfysize=\epsfxsize \multiply\epsfysize\epsftmp
       \multiply\epsftmp\epsftsize \advance\epsfrsize-\epsftmp
       \epsftmp=\epsfxsize
       \loop \advance\epsfrsize\epsfrsize \divide\epsftmp 2
       \ifnum \epsftmp>0
          \ifnum \epsfrsize<\epsftsize
          \else
             \advance\epsfrsize-\epsftsize \advance\epsfysize\epsftmp
          \fi
       \repeat
       \epsfrsize=0pt
     \else
       \epsfrsize=\epsfysize
     \fi
   \fi
}%
%
%
\def\epsfstatus#1{
   \ifepsfverbose
     \immediate\write16{#1: BoundingBox:
                  llx = \epsfllx\space lly = \epsflly\space
                  urx = \epsfurx\space ury = \epsfury\space}%
     \immediate\write16{#1: scaled width = \the\epsfxsize\space
                  scaled height = \the\epsfysize}%
   \fi
}%
%
%
{\catcode`\%=12 \global\let\epsfpercent=
\global\def\epsfatend{(atend)}%
%
%
%
%
%
%
%
\long\def\epsfaux#1#2:#3\\%
{%
   \def\testit{#2}
   \ifx#1\epsfpercent           
       \ifx\testit\epsfbblit    
            \epsfgrab #3 . . . \\%
            \ifx\epsfllx\epsfatend 
                \global\epsfatendtrue
            \else               
                \ifepsfatend    
                \else           
                    \epsffileokfalse
                \fi
                \global\epsfbbfoundtrue
            \fi
       \fi
   \fi
}%
%
%
\def\epsfempty{}%
\def\epsfgrab #1 #2 #3 #4 #5\\{%
   \global\def\epsfllx{#1}\ifx\epsfllx\epsfempty
      \epsfgrab #2 #3 #4 #5 .\\\else
   \global\def\epsflly{#2}%
   \global\def\epsfurx{#3}\global\def\epsfury{#4}\fi
}%
%
%
\def\epsfsize#1#2{\epsfxsize}%
%
%
\let\epsffile=\epsfbox
 